\documentclass[journal,comsoc]{IEEEtran}
\usepackage[T1]{fontenc}
\usepackage{cite}
\usepackage{amsmath,amssymb,amsfonts}
\usepackage{algorithm}
\usepackage{algpseudocode}
\usepackage{graphicx}
\usepackage{textcomp}
\usepackage{xcolor}
\usepackage{enumitem}
\usepackage{orcidlink}
\usepackage[acronym,toc]{glossaries}
\usepackage{color,soul}
\soulregister\gls{7}
\soulregister\glspl{7}

\usepackage[caption=false,font=normalsize,labelfont=sf,textfont=sf]{subfig}

\usepackage{nameref}
\newcounter{mylabelcounter}

\makeatletter
\newcommand{\labelText}[2]{%
#1\refstepcounter{mylabelcounter}%
\immediate\write\@auxout{%
  \string\newlabel{#2}{{1}{\thepage}{{\unexpanded{#1}}}{mylabelcounter.\number\value{mylabelcounter}}{}}%
}%
}
\makeatother

\newacronym{AoI}{AoI}{age of information}
\newacronym{AAoI}{AAoI}{average AoI}
\newacronym{PAoI}{PAoI}{peak AoI}
\newacronym{IoT}{IoT}{Internet of Things}
\newacronym{RTS}{RTS}{request to send}
\newacronym{CSMA}{CSMA}{carrier sensing multiple access}
\newacronym{BS}{BS}{base station}
\newacronym{MAM}{MAM}{max-age matching}
\newacronym{CSI}{CSI}{Channel state information}
\newacronym{EI}{EI}{energy information}
\newacronym{ED}{ED}{event-driven}
\newacronym{MDP}{MDP}{Markov decision process}
\newacronym{POMDP}{POMDP}{partially observable Markov decision process}
\newacronym{VoI}{VoI}{value-of-information}
\newacronym{MA}{MA}{mean availability}
\newacronym{MSE}{MSE}{mean square error}
\newacronym{MTTF}{MTTF}{mean time to failure}
\newacronym{IoTD}{IoTD}{\glsentryshort{IoT} device}
\newacronym{QoS}{QoS}{quality of service}
\newacronym{MAC}{MAC}{medium access control}
\newacronym{MTTR}{MTTR}{mean time to repair}
\newacronym{EH}{EH}{energy harvesting}
\newacronym{AIoT}{AIoT}{ambient \glsentryshort{IoT}}
\newacronym{ML}{ML}{machine learning}
\newacronym{EdN}{EdN}{edge node}
\newacronym{SoC}{SoC}{system-on-chip}
\newacronym{RL}{RL}{reinforcement learning}
\newacronym{IIoT}{IIoT}{industrial \glsentryshort{IoT}}
\newacronym{LoRAIN}{LoRAIN}{LoRa industrial network}
\newacronym{PHY}{PHY}{physical-layer}
\newacronym{WuS}{WuS}{wake-up signals}
\newacronym{WuR}{WuR}{wake-up radio}
\newacronym{TTI}{TTI}{transmission time interval}
\newacronym{knn}{KNN}{K-nearest neighbor}
\newacronym{dhc}{DHC}{device-hardware context}

\ifCLASSINFOpdf
\else
\fi

\usepackage{amsmath}
\usepackage[cmintegrals]{newtxmath}

\begin{document}
%
\title{Context-awareness for Dependable Low-Power IoT}
%
%
%

\author{David~E.~Ru\'{i}z-Guirola,~\IEEEmembership{Graduate Student Member, IEEE}, Prasoon~Raghuwanshi,~\IEEEmembership{Student~Member,~IEEE},
Gabriel~M.~de~Jesus, \IEEEmembership{Graduate Student Member, IEEE}, 
Mateen~Ashraf, 
and Onel L. A. L\'{o}pez, \IEEEmembership{Senior Member, IEEE}
\thanks{David E. Ru\'{i}z-Guirola, Prasoon Raghuwanshi, Gabriel~Martins~de~Jesus, Mateen Ashraf,  and Onel L. A. L\'{o}pez are with the Centre for Wireless Communications, University of Oulu, Finland. \{David.RuizGuirola, Prasoon.Raghuwanshi, Gabriel.MartinsdeJesus, Mateen.Ashraf, Onel.AlcarazLopez\}@oulu.fi.}
\thanks{This work has been partially supported by the Research Council of Finland (Grants 369116 (6G Flagship Programme) and 362782 (ECO-LITE)), the Finnish Foundation for Technology Promotion, and the European Commission through the Horizon Europe/JU SNS project AMBIENT-6G (Grant 101192113).}}

%
%

\markboth{Journal of \LaTeX\ Class Files,~Vol.~14, No.~8, August~2015}%
{Shell \MakeLowercase{\textit{et al.}}: Context-awareness for Dependable low-power IoT}
%



\maketitle

\begin{abstract}
Dependability is the ability to consistently deliver trusted and uninterrupted service in the face of operational uncertainties. Ensuring dependable operation in large-scale, energy-constrained Internet of Things (IoT) deployments is as crucial as challenging, and calls for context-aware protocols {where context refers to situational or state information}. In this paper, we identify four critical context dimensions for IoT networks, namely energy status, information freshness, task relevance, and physical/medium conditions, and show how each one underpins core dependability attributes. Building on these insights, we propose a two-step protocol-design framework that incorporates operation-specific context fields. Through three representative use cases, we demonstrate how context awareness can significantly enhance system dependability while imposing only minimal control-plane overhead. 
\end{abstract}


%
\IEEEpeerreviewmaketitle


\section{Introduction}\label{sec.I}

\IEEEPARstart{T}{he} \gls{IoT} is a cornerstone of modern digital transformation, enabling the seamless integration of physical objects with the digital ecosystem~\cite{mouhim2025towards}. It interconnects diverse devices equipped with sensors, actuators, {computing}, and communication modules to collect, exchange, and act upon data across various environments. These \glspl{IoTD} monitor parameters such as temperature, pressure, and motion, among others, while transmitting the corresponding data to centralized or edge servers for processing and/or control~\cite{mahmood2021industrial}. 

As \gls{IoT} deployments scale and begin to underpin critical infrastructure,  \gls{QoS} {requirements become more stringent. 
For example,} emerging {\gls{IoT}} applications {with tight \gls{QoS} demands in terms of reliability, latency, and/or energy storage lifetime include} wearable remote health monitoring, critical process management/control, and smart-grid automation~\cite{mahmood2021industrial}. 
Moreover, \glspl{IoTD} are shifting from passive reporters to coordinated agents, increasing data value while complicating control and lifecycle management. {This calls for} resource-adaptive mechanisms and judicious, fit-for-purpose wireless choices.  


As \gls{IoT} {starts} connecting critical infrastructures and {supporting} life-supporting applications, dependability becomes a core requirement. Indeed, dependability is the ability to consistently deliver trusted and uninterrupted service despite operational uncertainties, and it is a combination of key attributes, {like}: \textbf{(i) availability}, as the probability that the system is operational and able to deliver services at a given moment; \textbf{(ii) reliability}, as the probability that the system performs its intended function without failure over a specific time interval; \textbf{(iii) maintainability}, as the ability of the system to detect, diagnose, and repair faults quickly; \textbf{(iv) resilience}, as the system capacity to withstand and recover from disruptions, whether due to faults, failures, or environmental challenges; and \textbf{(v) scalability}, as the ability to accommodate growth from thousands to millions of \glspl{IoTD} (horizontal) and to sustain higher, time-varying traffic on existing deployments (vertical) without unacceptable degradation in other performance indicators.  

In addition to dependability features, sustainability emerges as a key complement for dependable \gls{IoT}. Indeed, 
\gls{IoT} systems must reduce reliance on external power sources through \gls{EH}, enabling long-term energy neutrality and ensuring continuous operation. Achieving dependability and sustainability goals demands a holistic view that spans beyond the communication stack (from the \gls{PHY} to the application layer) to include hardware and energy supply, edge/cloud computing, and user/task context. Specially due to stringent constraints like limited energy, hardware circuitry, and strict real-time demands.  
Traditional \gls{IoT} systems, particularly in low-power settings, operate under predefined rules. As a result, they cannot dynamically adapt to evolving conditions, leading to inefficiencies in energy usage, degraded \gls{QoS}, and subsequently reduced system dependability. In practice, most \gls{IoT} platforms still treat data acquisition and transmission as isolated functions and rarely exploit contextual cues from their surroundings to adjust operational parameters in real-time~\cite{mouhim2025towards}.

Context-aware protocols are appealing to address stringent dependability requirements. 
Context-aware in \gls{IoT} refers to any information that can be used to characterize the current state of an entity, whether it is an \gls{IoTD}, user, task, or environment. 
These protocols can dynamically adapt to network conditions and application needs to manage sensing, communication, {computing}, and actuation efficiently. Specific protocol applications range from efficient scheduling, task execution deference, edge computing, resource prioritization, and model-predictive control to \gls{ML} integration~\cite{mahmood2021industrial}. 
For instance, {energy and task context awareness may allow} nodes to anticipate communication needs, postpone non-essential tasks, or switch to energy-saving modes~\cite{albreem2021green}. 
Consequently, the {\gls{IoT}} protocol stack {should embed} delay, energy consumption, throughput, and/or fault tolerance information aspects thoroughly, specially in highly heterogeneous environments.

{Although context awareness has been studied in some IoT settings}~\cite{mouhim2025towards,10884945,Onel2025}, {a holistic, system-level integration tailored to low-power IoT for dependable operation remains incomplete. Key gaps include real-time handling of dynamic context and explicit, context-aware specifications for modeling, updating context under strict power budgets, and a missing link to dependability attributes beyond typical ones. Most importantly, the methodological bridge from raw context to device-level decision-making (\textit{e.g.}, sensing, actuating, transmission policies) is still missing.}  
This paper proposes a context-aware approach to protocol design as a foundation for allowing dependable operation in low-power, energy-constrained \gls{IoT} networks. 
We present an edge-centric control plane that performs a cross-layer mapping from context to dependability targets, deriving per-device policies.  
Additionally, we introduce application-specific awareness mechanisms that continuously configure system parameters based on contextual factors such as energy availability, network dynamics, and task requirements. 
Finally, through an assessment of representative use cases, we demonstrate significant improvements in detection latency, accuracy, and overall availability, relative to context-agnostic baselines, all achieved with minimal control-plane overhead.  

The rest of the paper is organized as follows. Section~\ref{sec.II} introduces dependability through context awareness, and describes a two-step protocol that encodes hardware and application-specific context. Section~\ref{sec.IV} demonstrates the effectiveness of context-aware operation via representative use cases comparisons, while Section~\ref{sec.V} outlines the challenges and open research directions. 
Finally, we conclude the paper in Section~\ref{sec.VI}.


\section{Context Awareness}\label{sec.II}




\begin{figure*}[t!]
    \centering
    \includegraphics[width=\linewidth]{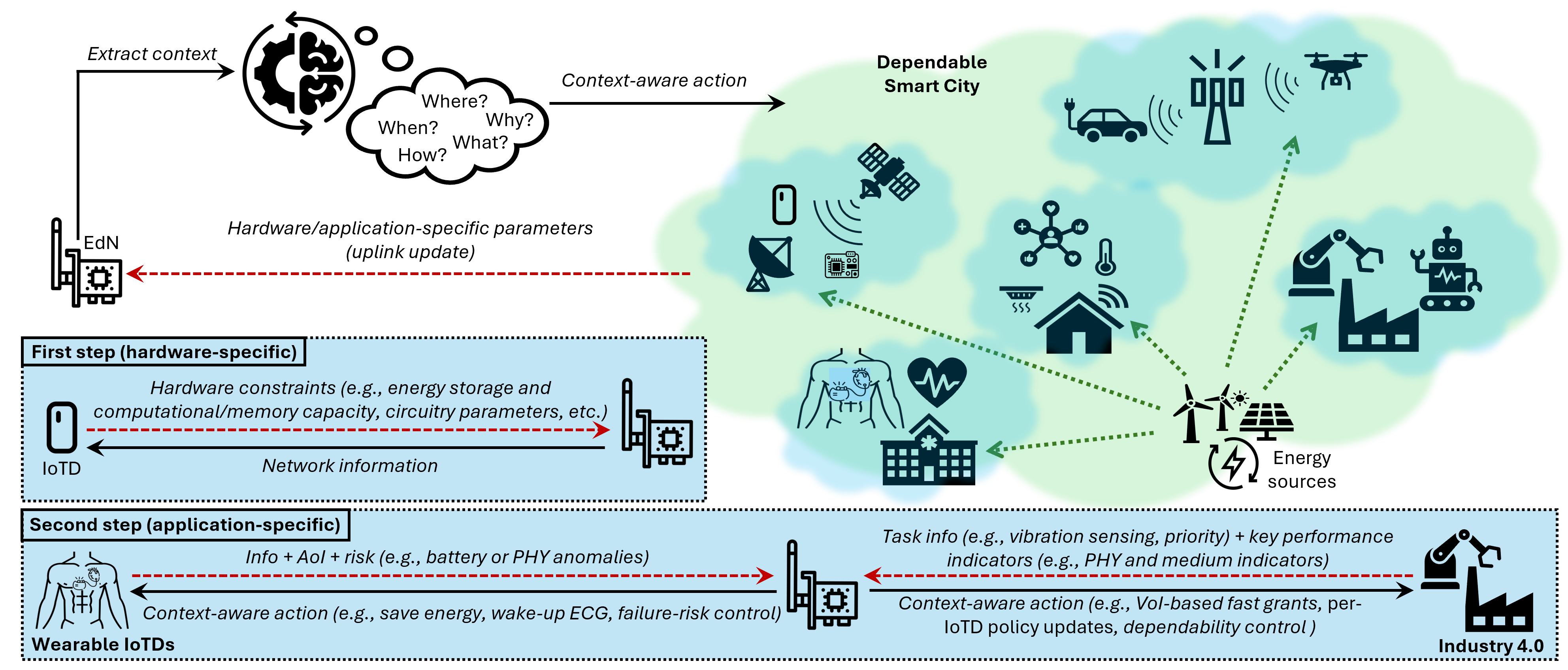}
    \caption{Illustrative example of context-awareness for dependable \gls{IoT}.}
    \label{fig:context_awareness}
\end{figure*}




\gls{IoT} context-awareness includes static attributes (\textit{e.g.}, device type, {hardware features,} or location), dynamic parameters (\textit{e.g.}, {environmental conditions} or remaining energy {in} storage), and behaviors (\textit{e.g.}, traffic patterns{, ongoing user/device activities} or system load). 
As shown in \figurename~\ref{fig:context_awareness}, context-awareness typically functions as an intermediary layer that collects system data, processes it, extract semantics, and supplies the resulting context to the control/protocol stack, facilitating better higher-level decision-making. By separating context acquisition from application logic, context-aware solutions relieve the system's burden of real-time context processing while preserving flexibility and scalability~\cite{albreem2021green}.




While context can encompass a broad spectrum of information, not all types are equally critical for dependable operation in low-power \gls{IoT} scenarios. Based on existing literature and system-level needs, this paper focuses on four context dimensions particularly relevant to enabling dependable low-power {\gls{IoT}} systems:
\begin{itemize}
    \item \textbf{Energy Awareness:} {Since} \glspl{IoTD} often operate under strict energy constraints, awareness of remaining energy in storage, charging opportunities, or \gls{EH} conditions enables intelligent scheduling and resource conservation~\cite{albreem2021green}.
    \item \textbf{Freshness-of-Information Awareness:} The value of information depends not only on accuracy but also on its timeliness. Awareness of \gls{AoI} helps in prioritizing updates and optimizing data flows~\cite{Chen:2022:ITIT}.
    \item \textbf{Task and Information Awareness:} The relevance and urgency of the data {that \glspl{IoTD} nodes} sense, transmit, or {process may vary dynamically}. {Being aware of the ongoing tasks' context can help} dynamically tailoring processing pipelines and offloading decisions~\cite{10884945}.
    \item \textbf{Physical and Medium Awareness:} Awareness of environmental, {\gls{dhc}}, and \gls{PHY} conditions, such as channel quality, interference, and node status, enables fault avoidance, reliable communication, and physical reconfiguration~\cite{sefati2024comprehensive}. 
\end{itemize}

Another important consideration in \gls{IoT} network design lies in where context is processed. Indeed, the cross-layer context may be aggregated and acted upon by a coordinator (\textit{e.g.,} in the cloud/edge) or processed locally at the nodes.
Centralized controllers globally optimize scheduling while providing strong reliability guarantees, but often introduce control overhead and slower fault recovery. In contrast, local context processing supports faster responses and better scalability, yet potentially deviating significantly from optimal policies due to limited visibility into the network state. In this work, we focus on an edge-based approach combining the reliability of centralized control with the responsiveness and scalability of local decision-making. 
{\textit{This requires an efficient protocol that encodes and subsequently conveys the underlying context-specific parameters among network entities.}} 

{A two-step procedure can be used to provide any type of context awareness,} as shown in \figurename~\ref{fig:context_awareness}. Specifically, in the \textit{first step} all those parameters that are related to the \textit{device's hardware constraints} and that can affect all the other types of contexts are encoded. These parameters can include, among others, energy storage capacity, processor computational capacity, transmitter/receiver circuitry parameters, memory size, interfaces/sensors, etc. {Since such \gls{dhc} parameters} do not change during the life cycle of \gls{IoTD}, they can be exchanged with the network only once during the registration phase. In the \textit{second step}, all the \textit{application-specific} parameters that can affect particular contextual performance are encoded in the protocol. The different types of context-specific parameters that the proposed approach uses in different context awareness are discussed in the following.



\subsection{Energy-awareness}\label{sec.III.1}

{A dependable energy-aware \gls{IoT} protocol must support resilience against power intermittency, adapting to fluctuating energy availability. There are different types of} \gls{EI} {that can be exploited, and these are related to the EH, stored energy, and/or load energy consumption processes. The cost/overhead for \gls{EI} acquisition depends heavily on the type, frequency, measurement point, and method} \cite{Onel2025, Sandhu_2021}. {In general, \gls{EI} granularity} can be broadly categorized into high-level and low-level.

High-level energy awareness in \glspl{IoTD} involves capturing coarse yet meaningful insights into energy availability and dynamics, enabling broader decision-making without deep task-specific introspection. Three essential components contribute to this abstraction. First, the energy measurement point specifies where the energy is monitored, whether at the energy transducer output (applicable in case of \gls{EH} \glspl{IoTD}), the energy storage element, or the load consumption point. Second, the energy measurement method defines how energy is quantified, which may include comparator-based techniques, information sampling, accumulation approaches, or indirect methods; each offering different trade-offs in terms of accuracy, complexity, and cost \cite{Onel2025}. Finally, energy forecasting involves predicting future energy availability based on historical or contextual data, using methods such as statistical modeling, heuristic-based prediction, or lightweight \gls{ML} approaches such as TinyML.

Low-level energy awareness involves fine-grained information about task-specific energy influencing parameters in the fundamental operations of \glspl{IoTD}. In sensing, parameters such as the sensing frequency, modality, and sensing signal voltage levels can be dynamically adjusted to balance energy usage with sensing accuracy. For computation operations, precision levels, inference types, and techniques like dynamic checkpointing, dynamic frequency scaling, and task scheduling influence how computational tasks adapt to energy availability. Communication tasks can be tuned through \gls{PHY} parameters, including transmit power levels, modulation schemes, and packet size, and through \gls{MAC} mechanisms such as duty cycling and the use of wake-up radios. Finally, actuation can be modulated by varying actuation intensity or by introducing physical and structural adaptations, enabling responsive but energy-aware interaction with the environment.




\subsection{Age of Information}\label{subseq:aoi}
{In cyberphysical} systems with important/critical {\gls{IoT}} monitoring components, status update messages are time-stamped, and the knowledge of both the content of the updates and the time they were measured or generated is required for proper operation \cite{Yates:2021:IJSAC}. In many of these systems, having the freshest possible information is critical, and the {\gls{AoI}}, defined as the time elapsed since the latest update from a given \gls{IoTD} was received, {can be used to quantify this. Typically} the \gls{AAoI} and \gls{PAoI} are considered as network-level \gls{AoI} performance metrics. 

{\gls{AoI} reduction may come} from design choices prioritizing other metrics or {from explicit AoI minimization frameworks}. In the latter, this is typically handled at the MAC layer by  somewhat favoring transmissions from \glspl{IoTD} with high \gls{AoI} \cite{Chen:2022:ITIT}. For instance, the \gls{EdN} can schedule the \glspl{IoTD} based on their \gls{AoI} and channel state, but \glspl{IoTD} deciding by themselves which will transmit is also possible. The latter may decrease exchanges between \gls{EdN} and \glspl{IoTD}, as well as energy consumption, but possibly affecting \gls{AoI} performance. Regardless of the approach's nature, effective scheduling requires the \gls{EdN} to know how \glspl{IoTD} operate, that is,  whether sensing is passive, event-driven, or periodic. For example, centralized scheduling may require fresh packets available at request, which is infeasible for event-driven sensing if no events happened recently. Thus, the \gls{EdN} must be aware of the characteristics of the packets generation. Such metadata must be shared when joining the network and updated if sensing conditions change, {while packets' age in the buffer must be regularly reported.}

\subsection{Task-aware Communication}\label{task_aware_comm}

Task-aware communication is about transmitting only the data {that is} relevant to the receiver {to} successfully complete its {application tasks} \cite{Holm_2023_10143239}.
{Properly implemented, this allows reducing} network congestion and the likelihood of transmission failure, {and} boosts the \gls{IoT} network's uptime,
{thus, increasing} dependability in terms of \textit{reliability} and network \textit{availability}.

Task-aware communication requires consideration of  \textit{relevance of transmitted data to the task} and the \textit{physical time}.
{Indeed,} the environment, observed by \glspl{IoTD}, evolves with the physical time,
{while} the information received from the environment's observation becomes stale over time from the task's perspective.
{Metrics like latency,} query-blind \gls{AoI} \cite{Holm_2023_10143239}, or \gls{AoI}-at-query \cite{Holm_2023_10143239} only reckon with the time attribute, not the relevance attribute.
Meanwhile, the \textit{pragmatic \gls{VoI}} metric takes into account both the physical time and transmitted data as well as its out-turn on the task \cite{Holm_2023_10143239}.
For example, consider {the system illustrated in Fig}.~\ref{fig:setup} where (i) \glspl{IoTD} observe {the state of} a dynamic process, (ii) the \gls{EdN} polls \glspl{IoTD} to gather information {about such state}, and (iii) external servers ask queries about {the current} {state} to the \gls{EdN}.
In this system, if a state estimator of the observed dynamic process is available at the \gls{EdN}, as in \cite{Holm_2023_10143239}, then the \textit{pragmatic \gls{VoI}} of the transmitted data is defined as the difference the transmitted data would make in the error in query response operation, {and the \gls{EdN} would devise a scheduling policy that minimizes the error in the query response.}


\subsection{\gls{dhc}-awareness and edge-computing  protocols}\label{sec:PHY}

By continuously monitoring \gls{IoTD}-level health indicators, {such as channel quality, interference levels, and node status (\textit{e.g.}, temperature, moisture ingress, or self-test/fault flags)}, systems can proactively detect early signs of degradation or failure. Recent studies have demonstrated the practical advantages of embedding \gls{dhc} awareness directly into \gls{IoT} architectures. For instance, {\gls{dhc}-aware} \glspl{SoC} {may utilize} runtime monitoring of local parameters, such as workload and temperature, to provide dependability by dynamically adjusting operation~\cite{sadiqbatcha2021real}. Similarly, \gls{RL}-based adaptive routing {may exploit} \gls{PHY}-derived link context with real-time feedback to strengthen paths and reduce retransmissions under dynamic conditions~\cite{ergun2022reinforcement}. 

Predictive resource allocation that leverages \glspl{IoTD} location and mobility can pre-position capacity where demand is likely, improving energy efficiency. Coupled with \gls{dhc}-level awareness and robustness features, the network can adapt proactively {by rerouting traffic away from vulnerable nodes, anticipating and avoiding bottlenecks, and triggering timely maintenance.} Periodic assessment of network health (\textit{e.g.,} \glspl{IoTD} health status and interference levels) can also trigger targeted actions such as node redistribution or route updates to maintain reliable and efficient operation. {When fused with semantic models or \gls{ML}, such awareness features further sharpen decision-making.} 

\gls{CSI} is one important \gls{PHY} context to enhance link reliability through smart resource allocation and acknowledgment schemes based on channel quality. 
Under tight energy budgets, it is important to combine budgeted pilot-aided \gls{CSI} estimation with non-coherent strategies, {which are much lighter.} These can help support robust link adaptation techniques, including outage-constrained selection and worst-case beamforming~\cite{lancho2023cell}. 
For example, the \gls{LoRAIN} protocol presents a centralized coordination with local fault tolerance, allowing \glspl{EdN} to switch channels, adjust time slots, or re-route based on real-time link conditions. 

The nature of \gls{IoT} tasks varies widely in terms of energy requirements for different applications, which must be carefully considered in protocol design. {For instance, a simple \gls{EI} acquisition protocol can be that where the \gls{IoTD} and \gls{EdN} are} capable of performing computing and communication tasks. {Two energy consumption lookup tables can be maintained in both \gls{IoTD} and \gls{EdN}: one for processor speeds and another for transmit powers. To represent these values, two multi-bit fields can be defined, with one encoding processor speed and the other the transmit power. The \gls{IoTD} can then specify its selected operating point by setting the corresponding bits in each field. This protocol design principle can likewise be extended to other fundamental \gls{IoT} tasks.}




\section{Context-Aware Protocols}\label{sec.IV}


\begin{figure}
    \centering
    \includegraphics[width=\linewidth]{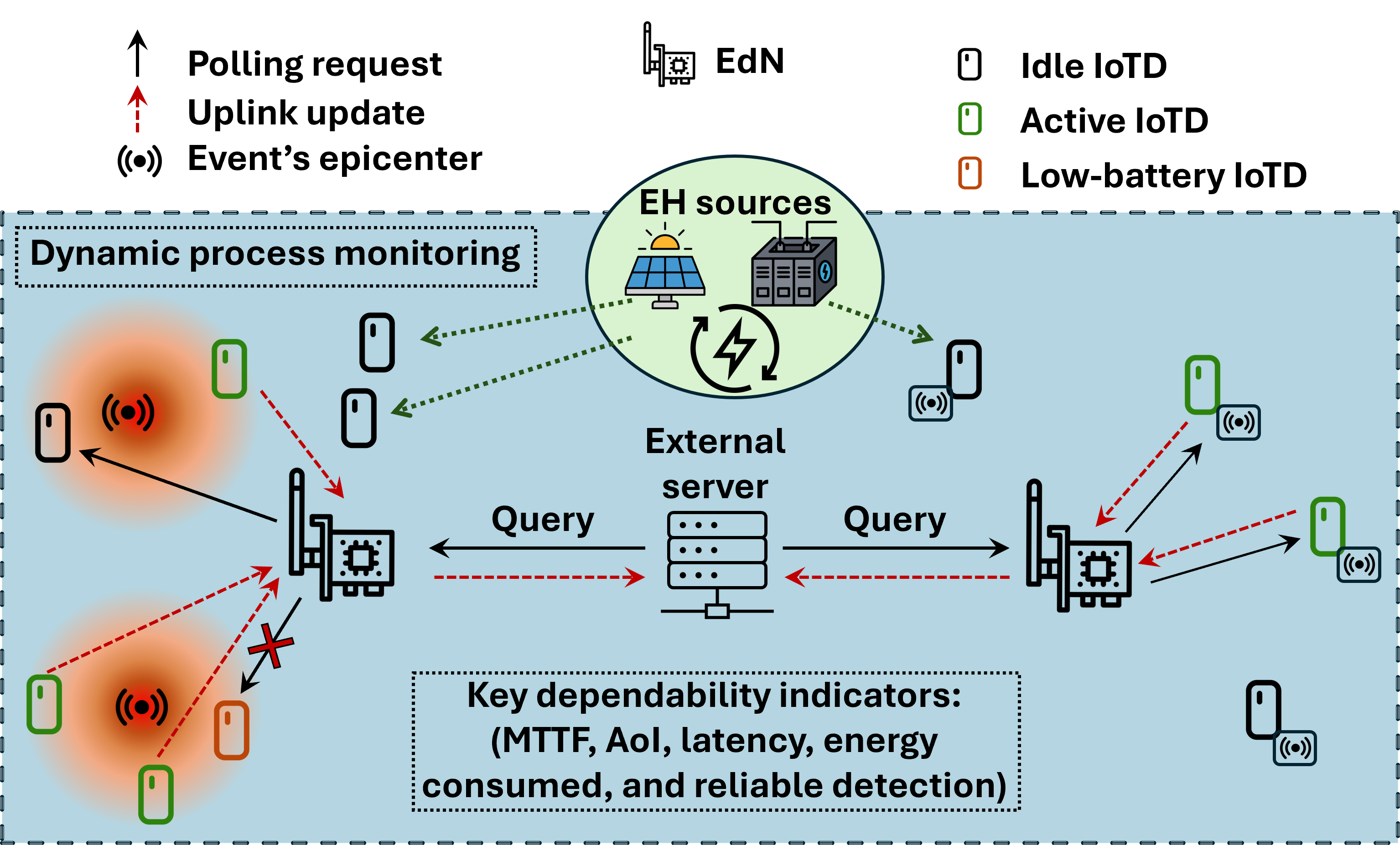}
    \caption{Energy-limited/\gls{EH}-powered network where the \glspl{EdN} coordinate dynamic processes monitoring such as event-driven activations (left) and uplink updates (right) from $N$ \glspl{IoTD}. The red circles indicate the full detection area for a given event, while the red cross indicates failed polling/uplink update.}
    \label{fig:setup}
\end{figure}
{Herein, we showcase how embedding context awareness improves dependability in low-power \gls{IoT} systems. 
We adopt the setup depicted in \figurename~\ref{fig:setup}, which underpins a variety of critical‐task scenarios from smart healthcare to automated homes, industrial deployments, and adaptive networks with non-stationary \glspl{IoTD} (cf.  \figurename~\ref{fig:context_awareness}). 
The \glspl{IoTD} are EH-powered for sensing and transmitting, while \glspl{EdN} query specific task information (e.g., vibration features, temperature, or CO$_2$ levels). To meet dependability guarantees under tight energy budgets and in the presence of harsh industrial interference, the system leverages context awareness along the four key dimensions discussed earlier. First, energy awareness allows \glspl{IoTD} to perform intelligent sensing and save energy without sacrificing critical event detection. Second, task awareness ensures that urgent alerts (e.g., motor overheat) preempt routine logging, so that scarce spectrum and processing resources are devoted to what matters the most. Third, information freshness guarantees that control decisions and predictive-maintenance queries are always based on the latest available data. Finally, medium awareness incorporates dynamic correlation collected by the edge, which continuously adapt transmission schedules to ensure full event detection. 
The following examples highlight how integrating context awareness in such setups can enhance dependability. 
The focus here is on context parameters that may change frequently during the network lifetime and hence are embedded in the second protocol step, and not on more rigid context parameters of the first protocol step.

\subsection{Critical Event Detection}\label{subsection_ED_WuR}

Consider an event-driven scenario, wherein the  \glspl{IoTD} operate under duty cycling, getting active based on proximity to critical events and transmitting data to the \gls{EdN} upon detection. To support timely decisions (\textit{e.g.}, machine adjustments or temperature calibration in smart industry or home), the \gls{EdN} may issue \gls{WuS} to idle \glspl{IoTD} that are either closer to the event or have higher energy reserves. This is enabled by a low-power \gls{WuR} module in each \gls{IoTD} for always-on listening. The \gls{EdN} must then optimize duty cycling to balance energy availability, reduce event detection latency, and minimize the risk of miss-detections. 
{We assume that the \glspl{IoTD} consume 1\% {of its maximum on-board energy storage capacity} per transmission interval when performing computing and sensing, while 10\% is consumed when communicating with the \gls{EdN}. The consumption for using the \gls{WuR} is assumed to be 0.07\% per transmission interval. Each \gls{IoTD} transmission and polling request is assumed to take 1 transmission interval (1 ms).}  

\figurename~\ref{fig:ED_latency} compares system dependability in terms of availability, reliability, and scalability, under context-aware intelligent duty cycling and sensing against a baseline using standard blind duty cycling. 
Context here includes the spatial and time correlation between \glspl{IoTD} and events, as well as the {\gls{EI} received at the \gls{EdN}} (\textit{i.e.,} \gls{EH}, consumption profile, and energy storage state of the \glspl{IoTD}).
The proposed context-aware scheme combines \gls{RL} with a \gls{knn}-based model to dynamically adjust duty cycling and wake-up decisions.  
As shown in \figurename~\ref{fig:ED_latency}, the proposed mechanisms improve event detection reliability by up to 3 times while reducing the event reporting delay by up to 56\%. 

Context-aware duty cycling and sensing also increase the \gls{MTTF}, defined as the expected time to failure (\textit{i.e.,} the average operating time before the \gls{IoT} system experiences a failure). Herein, we define \gls{MTTF} as the point at which more than 10\%, 25\%, and 50\% of the network area is covered by energy-depleted \glspl{IoTD}. These 10/25/50\% levels serve as operational tiers (\textit{e.g.,} minor degradation, impaired service, and critical failure), and model escalating mitigation actions. 
As illustrated in \figurename~\ref{fig:ED_latency}, our approach increases the \gls{MTTF} by up to two orders of magnitude while illustrating improvements in availability. 
Our dependability assessment encompasses availability, reliability, {resilience}, energy sustainability, and event detection latency. Overall, the proposed context-aware approach maintains long-term system dependability through intelligent sensing and energy-efficient operation.

\begin{figure}
    \centering
    \includegraphics[width=\linewidth]{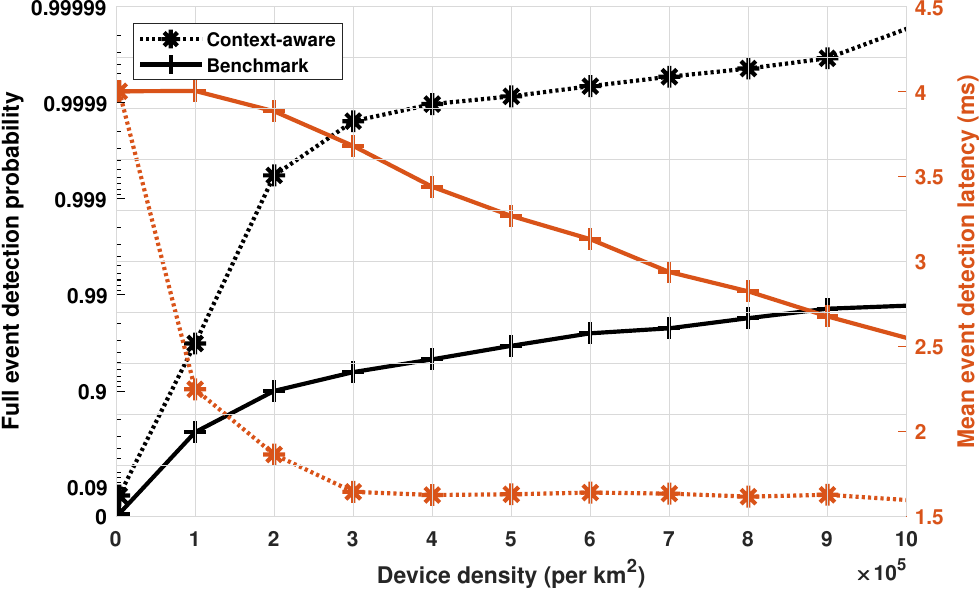}
    \begin{minipage}[t]{0.91\columnwidth}
    \includegraphics[width=0.985\linewidth]{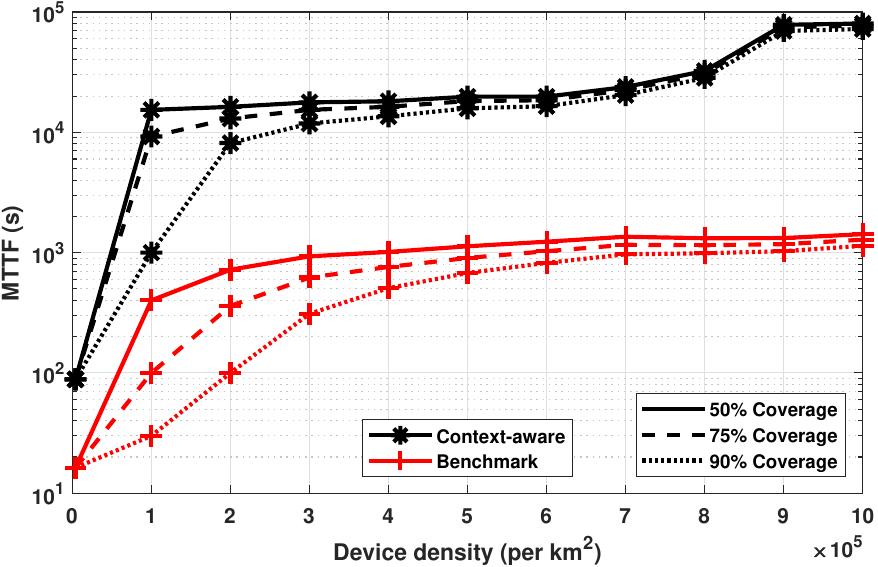}
    \end{minipage}\vfill
    \caption{Probability of mean event detection and event detection latency in ms (top), and \gls{MTTF} in seconds (bottom) for a context-aware approach and a context-agnostic benchmark as a function of the IoTD density. The benchmark relies solely on spatial correlation for wake-up requests, while duty-cycle parameters are fixed via brute-force tuning according to the event statistics.}
    \label{fig:ED_latency}
\end{figure}



\subsection{Wireless Networked Dynamic Process Monitoring}\label{subsection_WNDPM}

Consider a dynamic process monitoring system.
Here, the dynamic process state is represented in the form of a vector consisting of multiple state components, each observed by a single \gls{IoTD}.
The \gls{EdN} responds to queries asked by the external server(s) and aims to minimize the \gls{MSE} in the query response. 
A query is a request to send information about the particular characteristic of the dynamic process state. 
Moreover, the \gls{EdN} is equipped with a state estimator of the dynamic process and is not aware about the time steps at which queries would be asked.
Thus, in this system, the task-aware device scheduling problem can be modeled as \gls{POMDP}, wherein the \gls{EdN} acts as the \textit{agent} with \textit{action} space $\{0, 1, \cdots, N\}$, where \textit{action} $n \in \{1, \cdots, N\}$ means polling device $n$, and \textit{action} $0$ means refrain from polling \cite{raghuwanshi2024goal}.
The agent's \textit{state} is a function of
(i) the prior (before \gls{IoTD} transmission) estimates obtained from the state estimator,
(ii) the \gls{MSE} in query response with respect to the prior estimates, denoted as $\textrm{MSE}_{pri}$, and
(iii) the query-related parameter such as the time elapsed since the last query from an external server \cite{raghuwanshi2024goal}.
Moreover, when the agent polls an \gls{IoTD}, its \textit{reward} is the \textit{pragmatic \gls{VoI}} defined as the difference between $\textrm{MSE}_{pri}$ and \gls{MSE} in the query response with respect to the posterior (after \gls{IoTD} transmission) estimates \cite{raghuwanshi2024goal}.
Otherwise, the agent's \textit{reward} is $\textrm{MSE}_{pri}$ \cite{raghuwanshi2024goal}.
Here, the \gls{EdN} can exploit \gls{RL} to discover the device scheduling policy that solves the \gls{POMDP}, when it is not possible to derive its closed-form expression.

\begin{figure}[!t]
\captionsetup[subfigure]{labelformat=empty}
\centering
\begin{minipage}[t]{0.9\columnwidth}
\includegraphics[width=\linewidth]{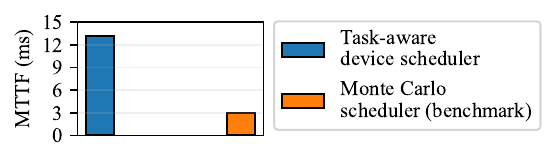}
\end{minipage}\vfill
\begin{minipage}[t]{0.48\columnwidth}
\includegraphics[width=\linewidth]{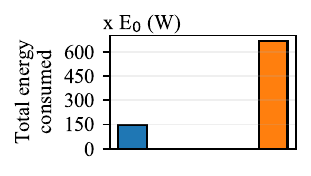}
\end{minipage}\hfill
\begin{minipage}[t]{0.52\columnwidth}
\includegraphics[width=\linewidth]{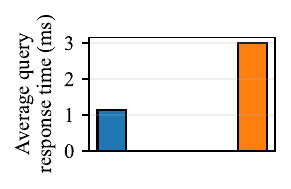}
\end{minipage}
\vspace{-5mm}
\caption{Illustration of \gls{MTTF}, total energy consumed at \glspl{IoTD}, and average query response time during the test run of the task-aware device scheduler and the Monte Carlo scheduler from \cite{raghuwanshi2024goal}.
Here, E$_0$ is the energy consumed, per transmission, at the device.
Moreover, the latency of the device-to-edge, edge-to-device, and edge-to-server transmission is $1$~ms.}
\label{TAC_monitoring_result}
\end{figure}

Let $t_{j, j-1}$ denote the time duration between the $j^{th}$ and $(j-1)^{th}$ polling instance from the task-aware device scheduler.
Note that the \gls{EdN} utilizes the prior estimates to respond to queries asked within the $t_{j, j-1}$ period.
Now, consider a situation where the network is down, i.e., the \gls{EdN} cannot poll \glspl{IoTD}, and external server(s) have asked queries to the \gls{EdN} during this time period.
Here the \gls{EdN} can continue responding to queries up to a time duration of $t_{j, j-1}$ even when the network is down.

For simulation results, consider a non-linear dynamic process, non-linear observation model on \gls{IoTD}, cubature quadrature Kalman filter as state estimator on the \gls{EdN}, and a \gls{RL}-based task-aware device scheduler with the aforementioned \gls{POMDP} \textit{state}/\textit{action}/\textit{reward}.
Fig.~\ref{TAC_monitoring_result} illustrates that the task-aware device scheduler reduces both the average query response time and total energy consumed at \glspl{IoTD} compared to {a} Monte Carlo {benchmark} scheduler {that polls \glspl{IoTD} only when a query arrives}.
Moreover, Fig.~\ref{TAC_monitoring_result} illustrates \gls{MTTF}, defined here as the average $t_{j, j-1}$, is significantly better for the task-aware scheduler. 
{{Here, a polling event is captured as a failure from the \gls{EdN} perspective. Indeed, a failure entails that the \gls{EdN} cannot reply immediately to the server(s) with the needed information, but it has first to poll the \glspl{IoTD}, consuming valuable time and energy resources.}}
Overall, the task-aware device scheduler \cite{raghuwanshi2024goal} increases both the \textit{reliability} and \textit{availability}, the dependability attributes, of the dynamic process monitoring system.

\subsection{Scheduling for AoI reduction}

Consider a request-based activation network, wherein $N$ devices communicate with a \gls{EdN} over $F$ channels. {When an \gls{IoTD} monitors its process, it generates a new packet. This monitoring can be done }either with a given probability $p$, or when requested by the \gls{EdN}. For simplicity, the channel undergoes on-off fading. Specifically, an \gls{IoTD} has a connection to the \gls{EdN} with probability $1-\varepsilon$, in a given channel at a given time-slot, and does not with probability $\varepsilon$. Moreover, transmissions fail in a channel when there are collisions from two or more  non-erased packets. {Next,} we compare three approaches of \gls{AoI}-based scheduling, namely the optimal and autonomous scheduling and the {threshold-based}.

\subsubsection{Optimal scheduling}
The \gls{EdN} {ranks the \glspl{IoTD}, whose channel are not erased (hence, based on \gls{CSI})} by their \gls{AoI} and {chooses the ones} with the highest {values}. With multiple orthogonal resources, several \glspl{IoTD} may transmit concurrently; otherwise, only the one with the highest \gls{AoI} and {acceptable} channel conditions transmits. While optimal for \gls{AoI}, this method requires complex computations and extensive \gls{EdN}–\gls{IoTD} communication, which may be prohibitive to implement.

\subsubsection{Autonomous scheduling}
\glspl{IoTD} autonomously schedule {their} transmissions. When an \gls{IoTD} has a data packet, it schedules a \gls{RTS} preceded by a wait time inversely proportional to its \gls{AoI}. Additionally, each \gls{IoTD} delays their \gls{RTS} to avoid collisions between packets from \glspl{IoTD} with the same \gls{AoI}. If the channel remains idle, the \gls{RTS} is sent. This method reduces communication overhead but is suboptimal as slots can go unused.

\subsubsection{Threshold-based}
When a \gls{IoTD} generates a packet, it is transmitted if the \gls{AoI} is above a threshold selected by the \gls{EdN}. This approach prevents low \gls{AoI} \glspl{IoTD} transmissions, but does not avoid collisions between packets of high \gls{AoI} \glspl{IoTD}, as they do not sense the channels before transmissions.

In Fig. \ref{fig:resultAoI}, we plot the \gls{AAoI} of networks following the three scheduling approaches against $p$. For the threshold-based, the age threshold is selected for each $p$ to provide the lowest possible \gls{AAoI}, {while} $p$ is irrelevant {for the centralized scheduling,} as \glspl{IoTD} generate their packets upon the \gls{EdN} request. Meanwhile, for the autonomous approach and the threshold-based, $p$ significantly influences the \gls{AoI} performance, as it leads to more \glspl{IoTD} contending for the transmission slots, therefore more packets being transmitted overall. The autonomous scheduling approaches the performance of the optimal centralized scheduling, but requires much less information exchange between \gls{EdN} and \glspl{IoTD}. On the other hand, the threshold-based has its best performance at $p=0.4$, but is still far from the optimal. The choice between these scheduling approaches depends on the network's characteristics, but ultimately on the \glspl{IoTD}' capabilities. If packets are generated often enough, opting for the autonomous approach is more advantageous, while the optimal scheduling is a good choice when packets can be generated at will and the performance requirements are stringent, as long as the \gls{EdN} and \glspl{IoTD} can afford the overhead. If communication between entities is not possible, the threshold-based strategy can be employed.  

\begin{figure}
    \centering
    \includegraphics[width=0.9\linewidth]{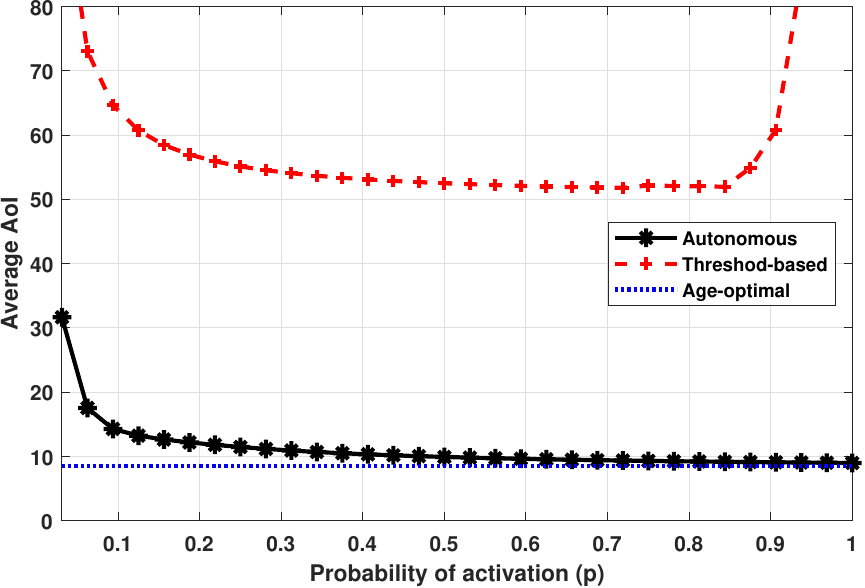}
    \caption{Comparison of three AoI-based approaches {as function of the probability of activation $p$ for a network with $N=32$ \glspl{IoTD}, $F=2$ channels, and $\varepsilon=0.5$.} The optimal approach guarantees low \gls{AAoI} at the cost of increased communication between \glspl{IoTD} and \gls{EdN}. The autonomous scheduling also decreases the \gls{AAoI}, specially at higher values of $p$, but \gls{IoTD} are required to listen to the channel and communicate among themselves. The threshold-based achieves a higher \gls{AAoI}, but no signaling between any entities is required.}
    \label{fig:resultAoI}
\end{figure}





\section{Challenges and Open Research Directions}\label{sec.V}


One of the main challenges in \gls{IoT} protocol design arises due to the diversity of possible operations that an \gls{IoTD} is expected to perform. This introduces the difficulty of encompassing all the necessary information (operational/hardware parameters and operational requirements) within a unified protocol. Therefore, a possible future direction is to identify the major performance contributors and refine the protocol design based on this identification.


Designing context-aware protocols for \gls{IoT} networks becomes significantly more challenging when multiple contexts are simultaneously involved. Owing to the inherently multi-dimensional functioning of modern \glspl{IoTD}, protocols should allow heterogeneous, dynamic, and sometimes conflicting information exchange. This multi-facet awareness raises challenges regarding context fusion, conflict resolution, and scalability. Furthermore, the real-time nature of many \gls{IoT} applications demands that these conceived protocols perform context awareness with minimal/maximal latency/resource utilization, all while ensuring dependability. 

Leveraging context awareness often relies on an explicit mathematical modeling of the underlying dynamic process. However, obtaining a mathematical model for every dynamic process might not be possible. Without this model information, context-aware communication cannot be effectively implemented. Consequently, these systems would be unable to experience the respective dependability-based benefits that context-aware communication provides. Therefore, developing mechanisms that do not rely on mathematical models remains an open research challenge.


\section{Conclusion}\label{sec.VI}

We demonstrated that embedding four context dimensions (energy status, information freshness, task relevance, and channel conditions) enables low-power IoT devices to meet stringent dependability targets, including high availability, reliability, and timely delivery, even under severe resource constraints. We proposed a two-step field-based protocol design so that each form of awareness can be enabled as required.
By feeding context directly into sensing, scheduling, and transmission decisions, devices transmit only the data that each task demands, minimizing unnecessary updates and conserving energy. Three representative numerical examples demonstrated how this context-driven approach yields dependable, energy-efficient operation in dynamic, resource-limited environments. Finally, our results confirmed that context-aware protocol design provides a robust foundation for a dependable IoT stack, capable of continuous, trusted service in demanding scenarios.

\ifCLASSOPTIONcaptionsoff
  \newpage
\fi


\bibliographystyle{IEEEtran}
\bibliography{references_short}


\section*{Biographies}

\noindent\textbf{David E. Ruiz-Guirola} [S'19] is currently pursuing a Ph.D. degree focused on sustainable IoT, machine learning, and traffic prediction, at the University of Oulu, Finland.

\noindent\textbf{Gabriel Martins de Jesus} [S'25] is pursuing a Ph.D. degree in communications engineering, focusing on active users detection for IoT networks, at the University of Oulu, Finland.

\noindent\textbf{Prasoon Raghuwanshi} [S'23] is pursuing a Ph.D. degree in communications engineering, focusing on random access protocols for IoT networks, at the University of Oulu, Finland. 

\noindent\textbf{Mateen Ashraf} is working as a postdoc researcher at the University of Oulu, Finland. His research interests include protocol design for energy harvesting IoT devices.

\noindent\textbf{Onel L\'opez} [S'17, M'20, SM'24] is an Associate Professor of wireless communications engineering, focused on sustainable IoT connectivity, at 6G Flagship, University of Oulu, Finland.

\end{document}